\documentclass[aps,pre,showpacs,amsmath,amssymb,amsfonts,lengthcheck,twocolumn,longbibliography,superscriptaddress, nofootinbib]{revtex4-2}

\usepackage{changes} 
\usepackage{cancel}

\usepackage{graphicx}
\usepackage{subfigure}
\usepackage{amsthm}
\usepackage{amsmath}
\usepackage{verbatim}
\usepackage{dcolumn}
\usepackage{bm}
\usepackage{color}
\usepackage[colorlinks=true,citecolor=blue,linkcolor=blue,urlcolor=blue]{hyperref}%
\usepackage{xcolor}
\usepackage{dsfont}
\usepackage{longtable}
\usepackage{tabularx}
\usepackage[ruled,vlined]{algorithm2e}
\allowdisplaybreaks
\newcommand{\bra}[1]{\left\langle #1\right|}
\newcommand{\ket}[1]{\left|#1\right\rangle}

\newcommand{\tr}[1]{\mathrm{tr}\left\{#1\right\}}

\newcommand{\bla}{bla\\bla\\bla\\bla\\bla}

\begin{document}

\title{Unifying Collisional Models and the Monte Carlo Metropolis Method: Algorithms for Dynamics of Open Quantum Systems}
\date{\today}
\author{Nathan M. Myers}
\email{myersn1@vt.edu}
\affiliation{Department of Physics, Virginia Tech, Blacksburg, VA 24061, USA}
\author{Hrushikesh Sable}
\affiliation {Department of Physics, Virginia Tech, Blacksburg, VA 24061, USA}
\author{Vito W. Scarola}
\affiliation {Department of Physics, Virginia Tech, Blacksburg, VA 24061, USA}

\begin{abstract}
Collisional models, or repeated interaction schemes, are a category of microscopic open quantum system models that have seen growing use in studying quantum thermalization, in which the bath is modeled as a large ensemble of identical ancillas that sequentially interact with the system. We demonstrate an equivalence between the system dynamics generated by the collisional model framework and the Metropolis algorithm, subject to two primary conditions. Namely, that each collisional model bath ancilla is prepared in a thermal state with a discrete spectrum that matches the energy eigenstate transitions of the system, and that the ratio of the ancilla partition function to the number of system eigenstates remains small. If these conditions are satisfied, the system dynamics generated by both methods are identical for arbitrary initial states and in both the steady state and transient regimes. This allows the typically purely phenomenological Metropolis scheme to be used as a tool to study exact pre-thermalization dynamics without the need to explicitly model the system-bath interaction.    
\end{abstract}

\maketitle

\section{Introduction} 
\label{sec:1}

The mechanisms and conditions under which a many-body quantum system will thermalize is a question of significant interest that bridges the fields of quantum thermodynamics, condensed matter physics, atomic, molecular, and optical physics, and quantum information. In the context of open quantum systems it is well established that a quantum system coupled to a heat bath environment evolving under the Markovian Lindblad master equation will equilibrate to a thermal state at the temperature of the bath \cite{BreuerPetruccione2007}. While the Lindblad equation is often tractable, it relies on strong assumptions about the dynamics of both the system and environment as well as the system-environment coupling, namely the Born-Markov and rotating wave approximations \cite{BreuerPetruccione2007}.

In recent years another approach to modeling open quantum systems has seen growing use, especially in the field of quantum thermodynamics, known as collisional models, or repeated interaction schemes \cite{Ciccarello2022}. In the collisional model approach the environment is assumed to consist of a collection of many identical subsystems, referred to as ancillae. The interaction between the system and environment occurs as a series of discrete unitary interactions (``collisions") between the system and one ancilla of the environment.

By microscopically modeling the system-environment interaction, collisional models have proven particularly useful in studying non-Markovian dynamics \cite{Bernardes2014, Bernardes2015, Ciccarello2013, McCloskey2014, Kretschmer2016, Camasca2021, Saha2024}, non-equilibrium dynamics \cite{Karevski2009, Barra2015, Strasberg2017, Seah2019, Rodrigues2019, Guarnieri2020}, quantum thermometry \cite{Alves2022, Seah2019PRL}, quantum synchronization \cite{Karpat2019, Li2023}, strongly-correlated models \cite{Karevski2009, Andreys2020}, quantum batteries and thermal machines \cite{DeChiara2018, Landi2021, Melo2022}, quantum optics \cite{Ciccarello2017, Filippov2022}, and modeling noise in quantum devices \cite{Kretschmer2016}. Collisional models have also served as a successful framework for studying concepts at the intersection of information theory and thermodynamics such as information scrambling \cite{Li2020, Li2022}, Landauer's principle \cite{Lorenzo2015, Pezzutto2016}, and quantum Darwinism \cite{Campbell2019, GarciaPerez2020}. Notably, the typical Lindblad master equation can be derived from the collisional model framework under the assumptions of non-interacting, uncorrelated ancillae \cite{Ziman2005, Ciccarello2017, Cattaneo2021, Ciccarello2022}. 

The conditions necessary for a collisional model to result in thermalization has seen significant study \cite{Terhal2000, Scarani2002, Barra2017, Arisoy2019, Jacob2021, TabaneraBravo2023}. A critical component for achieving thermalization is the condition that the environment ancillae couple to each transition energy of the system \cite{Arisoy2019}. This is necessary to ensure that the system Hilbert space is fully explored and each energy eigenstate can be populated.

A notable drawback of the collisional model approach is the need to operate in the joint system-bath Hilbert space, which can become computationally unwieldy, especially for large-dimensional bath ancillae. Nevertheless, collisional models are conceptually important as they provide a microscopic framework that can operate outside of common assumptions such as weak system-bath interactions.

Distinct from the deterministic evolution in collisional models, Monte Carlo methods are another class of methods that rely upon stochastic sampling of states in the Hilbert space such that the dominant contributions to the ground or the thermal states are captured \cite{LandauBinder_2014, Walter2015}. These are iterative, Markovian algorithms wherein each iteration, a random change to the state is considered and either accepted or rejected based on a relative probability criterion. By stochastic averaging over different iterations, the underlying probability distribution can be built. The process must satisfy the principles of ergodicity and detailed balance, meaning that every state can be connected to any other state through a finite number of moves, to ensure a complete sampling of states. Such Monte Carlo techniques are utilized in understanding the thermodynamic properties of lattice systems \cite{Sandvik2010, Ceperley2003} and can also be interpreted as a dynamic process related to the Glauber Kinetic Ising models \cite{Glauber2004, Augusiak2010}.  

Here we perform a detailed comparison between a typical collisional model approach of periodic system-bath interactions and a modified Monte Carlo Metropolis algorithm. We show that a Metropolis Monte Carlo scheme can replicate collisional model dynamics in both the transient and steady state regimes when the Metropolis time step is chosen to be the same as the time between collisions in the collisional model. Our equivalence holds under the assumption of thermalizing dynamics generated by interaction with a bath whose spectrum couples to each of the system's energy eigenstate transitions. It is also subject to the condition that the ratio of the bath ancillae partition functions to the number of system energy eigenstate transitions is small, which itself depends on an interplay between temperature, model energy spectrum, and system size. 

Notably, by modifying the rejection step procedure to implement a decoherence process the Metropolis algorithm, like the collisional model, can generate dynamics from arbitrary initial states, including coherent states. This equivalence shows that the Metropolis algorithm can be used not only as a phenomenological model for reaching thermalization, but can, under some circumstances, fully capture the transient dynamics of thermalization. Furthermore, as the Metropolis algorithm can be carried out without directly modeling the bath, this equivalence allows for thermalizing collisional model dynamics to be generated while only working in the Hilbert space of the system. While both methods still require diagonalizing of the system Hamiltonian to generate the transitions between energy eigenstates, the Metropolis algorithm posses a computational memory advantage over the collisional model since it only needs to store matrices that scale with the system Hilbert space, rather than the larger joint system plus bath Hilbert space.

In sections \ref{sec:2} and \ref{sec:3} we introduce the collisional model and Metropolis algorithms, respectively, and demonstrate how both approaches can lead to thermalization. In section \ref{sec:4} we derive an analytical equivalence between the dynamics produced by both models, and numerically demonstrate the conditions under which this equivalence is achieved using the Heisenberg XXZ chain as a representative model.

\section{Collisional model thermalization}
\label{sec:2}

\begin{figure}
    \includegraphics[width=0.25\textwidth]{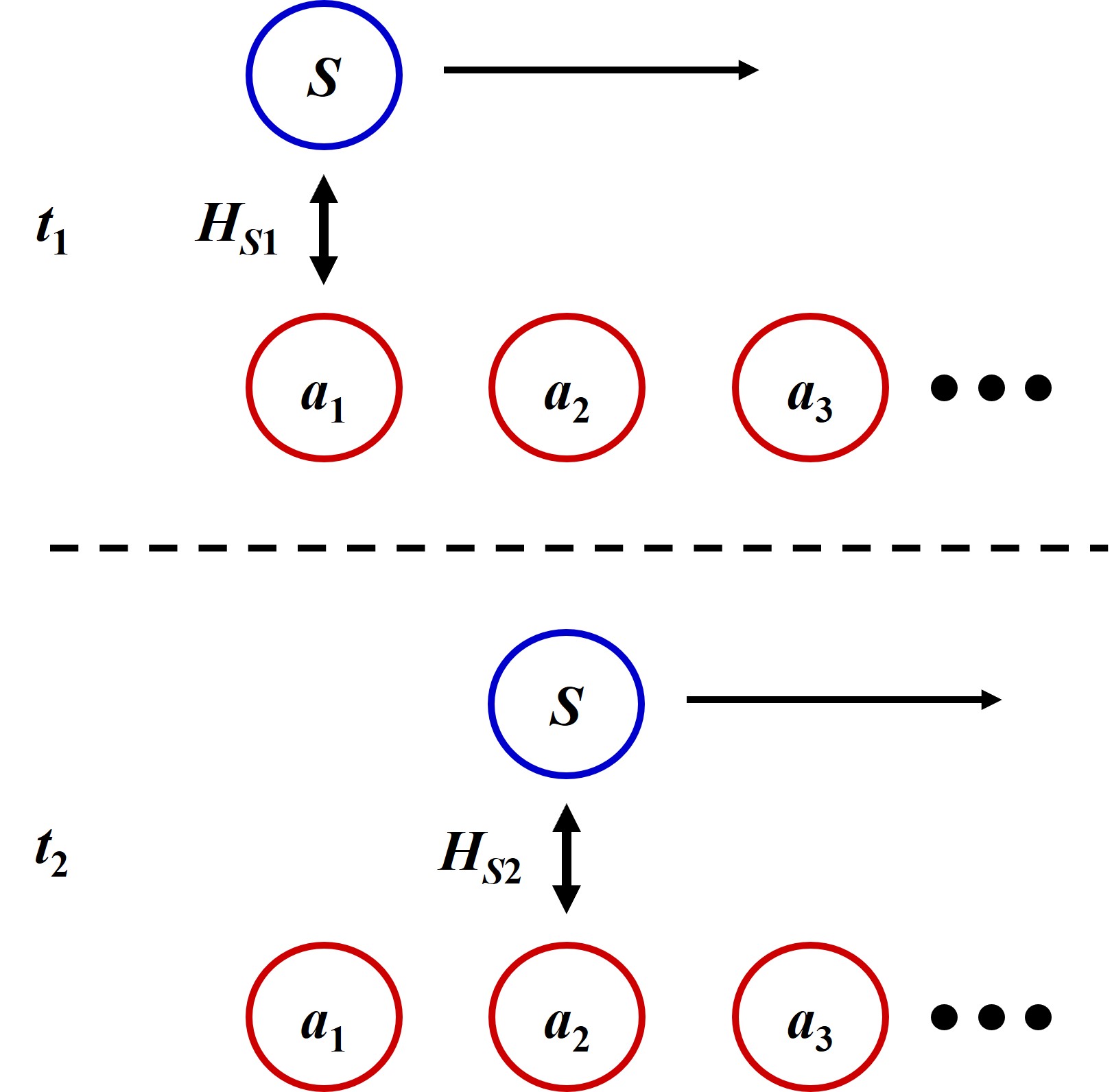}
    \caption{Illustration of the collision model approach. $S$ represents the system and $a_i$ labels bath ancilla $i$. The system-bath interaction $H_{S1}$ occurs at time $t_1$ whereas the interaction $H_{S2}$ occurs at a later time $t_2$.}
    \label{fig:CMsketch}
\end{figure}

Following the typical repeated interaction framework we consider a quantum system $S$ and a collection of non-interacting environment ancilla systems $a_j$. Initially, for a period of $t_S$ the system evolves freely under the dynamics generated by $H_S$. Then at time $t_1$ the system interacts with environment ancilla $a_1$ for time $\Delta t$, governed by the interaction Hamiltonian $H_{S1}$.  After the interaction, the environment is traced out, yielding system state $\rho_S(t_1)$,
\begin{equation}
    \label{eq:CMevolution}
   \rho_S(t_1) = \mathrm{tr}_{a_1}\{U_{S1}[U_S \rho_S(t_0) U_S^{\dagger} \otimes \rho_{a_1} ]U_{S1}^{\dagger}\} 
\end{equation}
This process, free evolution followed by interaction with a fresh environment ancilla, is then repeated for $n$ time steps. Note that the time interval between collisions $t_S = t_{n+1}-t_n$ depends on the physical properties of the system and environment, such as the system's scattering cross section and the density of particles (ancillae) in the environment. In Fig. \ref{fig:CMsketch} we provide a conceptual illustration of the collisional model approach.  

Ultimately, we are interested in how this process leads to the thermalization of the system, quantified by whether $\rho_S(t_n)$ approaches a Gibbs state at inverse temperature $\beta = 1/k_B T$, 
\begin{equation}
    \rho_S(t_n) \rightarrow e^{-\beta H_S}/Z_S.
\end{equation}
where $Z_S = \tr{e^{-\beta H_S}}$ is the partition function of the system. We begin by assuming that each bath ancilla is identical and initialized in a thermal state,
\begin{equation}
    \rho_{a} = e^{-\beta H_a}/Z_a
\end{equation}
Note that, in general, we should not expect the bath thermal state to be the same as that of the system. However, in order to achieve thermalization, the structure of the bath cannot be arbitrary either. 

The thermal state is diagonal in the energy eigenbasis, with the population of eigenstate $j$ given by the Boltzmann factor $e^{-\beta E_j}/Z_S$ where $E_j$ is the corresponding eigenenergy. For thermalization of an arbitrary initial state, in order to populate all the energy eigenstates, the system-bath interaction must couple each possible energy eigenstate transition in the system to a corresponding transition in the bath \cite{Arisoy2019}. For microscopically sized baths where the bath spectrum can be well approximated as continuous this condition is trivially satisfied. However, for a bath with a discrete spectrum, as is often the physically relevant case for many-body quantum thermalization, this transition energy matching condition is a crucial consideration.

\subsection{Collisional model thermalization of the XXZ model}

As a demonstrative example for how the collisional model can produce thermalization of a many-body quantum system, we consider an $N$-site one-dimensional XXZ model with open boundaries characterized by the Hamiltonian,
\begin{equation}
    H_{XXZ} = -J \sum_{q = 1}^{N-1} \left( \sigma^x_q \sigma^x_{q+1} + \sigma^y_q \sigma^y_{q+1} + \Delta \sigma^z_q \sigma^z_{q+1} \right) + \frac{h}{2} \sum_{q = 1}^{N} \sigma^z_q 
    \label{hamil_xxz}
\end{equation}
where $\sigma^{\alpha}$, $\alpha \in \{x,y,z\}$ are the Pauli matrices. First, let us consider the simple case of $N=2$. In this case the eigenenergies are,
\begin{equation}
\begin{split}
    &E_1 = -h-J\Delta, \quad \, E_2 = J(\Delta-2), \\ &E_3 = h-J\Delta, \qquad E_4 = J(\Delta+2).
\end{split}
\end{equation}

Let us now consider the structure of the bath. In order to guarantee the energy matching condition for thermalization is fulfilled we assume that the bath consists of an $M+1$-level system where $M$ is the total number of transitions between the system's energy eigenstates. The bath's energy levels are spaced such that each transition energy in the system, $E_i - E_j$, corresponds to an energy gap between the bath ground state and a corresponding excited state $\ket{\alpha_{i,j}}$, $\epsilon_{\alpha_{i,j}} - \epsilon_{0}$. The thermal state of each bath ancilla is thus,
\begin{equation}
    \label{eq:BathTherm}
    \rho_a = \frac{1}{Z_a} \sum_{\alpha_{i,j} = 0}^{M} e^{-\beta \epsilon_{\alpha_{i,j}}} \ket{\alpha_{i,j}}\bra{\alpha_{i,j}} 
\end{equation}
where $Z_a = \sum_{\alpha_{i,j}} e^{-\beta \epsilon_{\alpha_{i,j}}}$ is the typical partition function. The bath operators, $B_{\alpha_{i,j}} \equiv \ket{0}\bra{\alpha_{i,j}}$ correspond to the jump operator between the ground state of the bath and excited state $\ket{\alpha_{i,j}}$. Similarly, the system operators, $A_{i,j} \equiv \ket{j}\bra{i}$ correspond to the jump operator between the system energy eigenstates. Assuming that the system energy eigenvalues are labeled such that $E_1 \leq E_2 \leq E_3 ...$, the interaction Hamiltonian can be written as, 
\begin{equation}
    \label{eq:InteractionH}
    H_I = g \sum_{i>j} \left(\ket{j}\bra{i} \otimes \ket{0}\bra{\alpha_{i,j}} + \ket{i}\bra{j} \otimes \ket{\alpha_{i,j}}\bra{0} \right)
\end{equation}  

\begin{figure}
	\subfigure[]{
            \includegraphics[width=.22\textwidth]{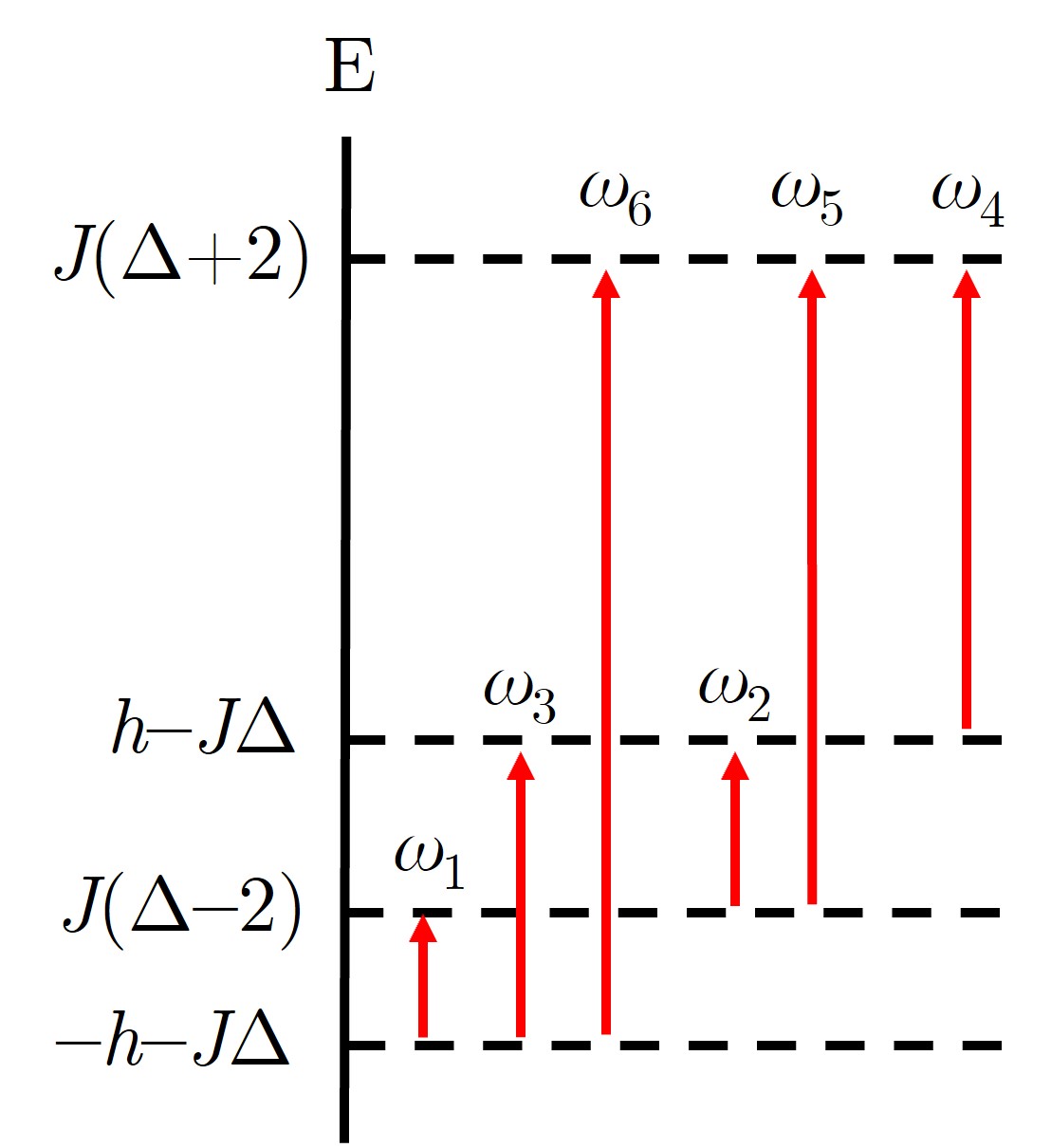}
	}
	\subfigure[]{
		\includegraphics[width=.18\textwidth]{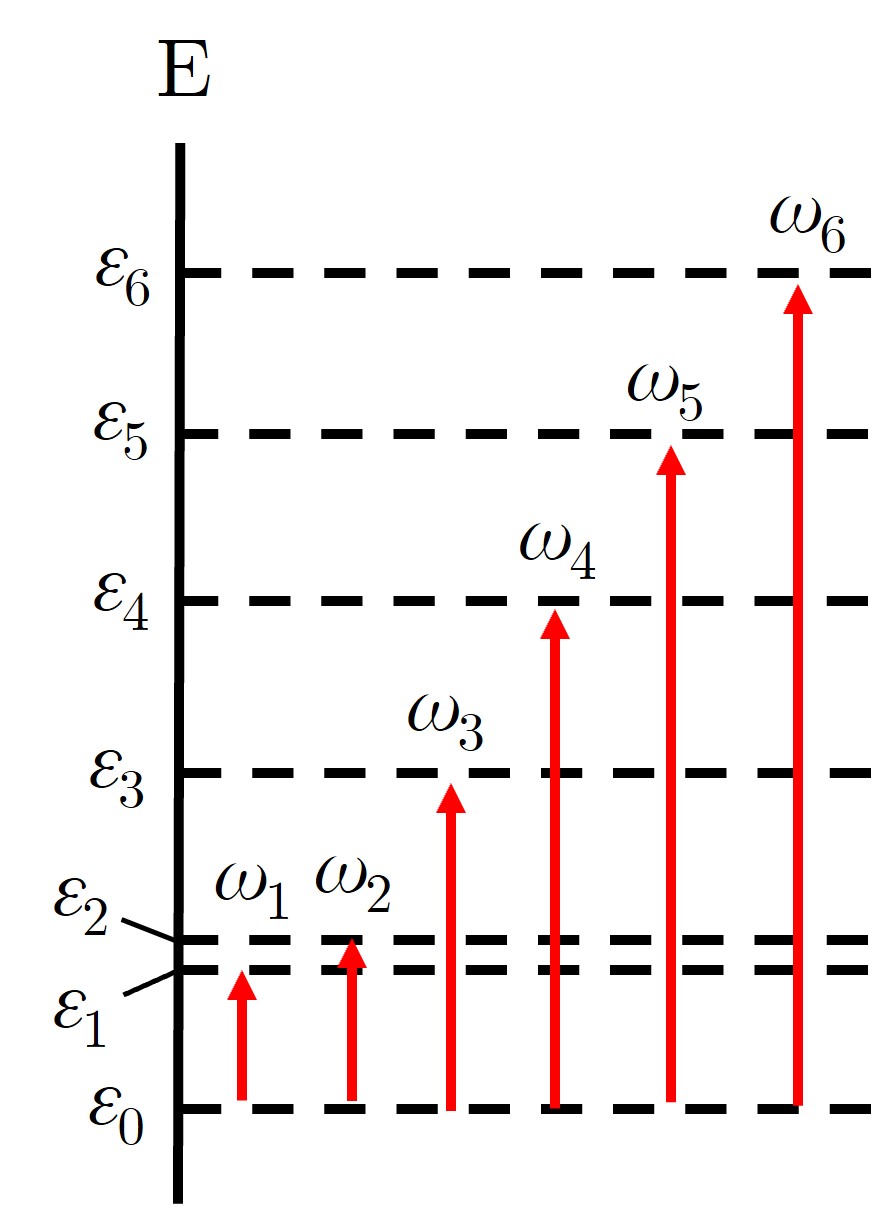}
	}
	\caption{\label{fig:SpectraMatch} Diagrams of the (a) system and (b) bath energy spectra for the one-dimensional two-site XXZ model. The red arrows indicate all possible transition energies in the system, and show how those transitions are matched in the bath spectrum.}
\end{figure}

For the XXZ chain with $N=2$, there are $M = \binom{2^N}{2} = 6$ possible system energy eigenstate transitions. Thus, in this case, each bath ancilla has a dimension of $\binom{2^N}{2} + 1$. The spectra for both the system and the bath ancillae is plotted in Fig. \ref{fig:SpectraMatch}, demonstrating how the bath spectrum fulfills the transition energy matching condition. 

We note that this is only one possible approach to constructing a bath that fulfills the energy matching condition. For example, another approach using low-dimensional ancillae with time-dependent energy gaps that are rapidly swept across the full system energy spectrum has also been demonstrated to lead to thermalization \cite{Metcalf2020}.

The free evolution for the system is generated by the unitary operator,
\begin{equation}
\label{eq:freeU}
    U_S = e^{- i H_{XXZ} t_S}
\end{equation}
while the interaction unitary for each system-ancilla interaction is,
\begin{equation}
\label{eq:IntU}
    U_{Sa} = e^{- i H_I \Delta t}.
\end{equation}

\begin{figure}
    \includegraphics[width=0.4\textwidth]{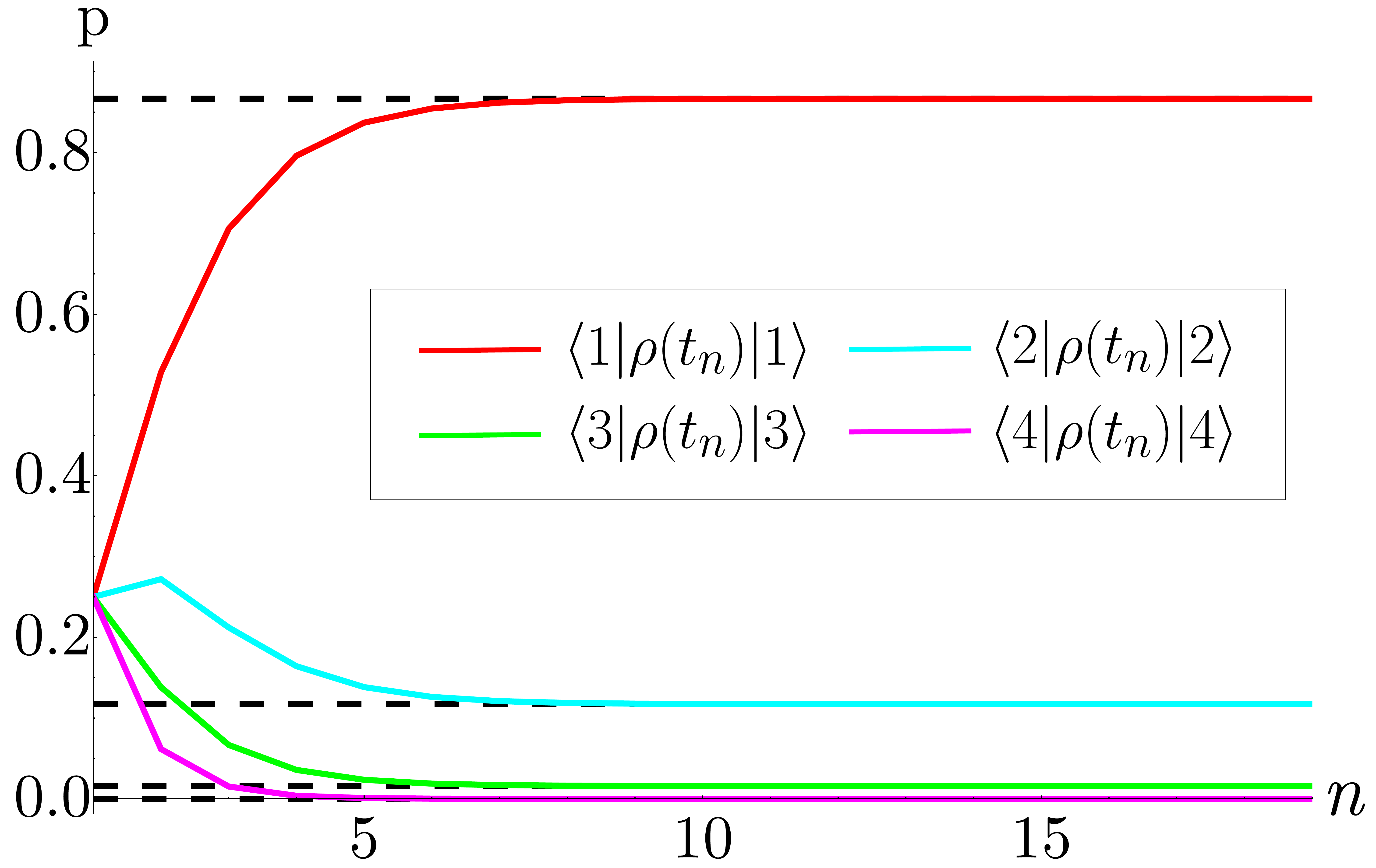}
    \caption{Occupation probabilities of each energy eigenstate of the two-site XXZ model as a function of the collisional model time step. Dashed horizontal lines indicate the thermal state occupation probabilities for each eigenstate. Parameters are $J = h = \Delta = 1$, $t_s = \Delta t = 1$, $g = 1$, and $\beta = 2$.}
    \label{fig:ThermXXZCM}
\end{figure}

Combining Eqs.~\eqref{eq:freeU}, \eqref{eq:IntU}, and \eqref{eq:BathTherm} in Eq.~\eqref{eq:CMevolution}, we numerically simulate a 20 time step collisional model for the two-site XXZ chain. We assume the initial system state to be a pure state composed of an equal superposition of all energy eigenstates $\ket{\psi_0} = 2^{-N/2}\sum_i \ket{i}$. We note there is nothing special about this choice of initial state, and thermalization will occur from an arbitrary initial state. In Fig. \ref{fig:ThermXXZCM} we plot the occupation probabilities of each energy eigenstate as a function of the time step, $n$. We see that the occupation probabilities rapidly approach the thermal state values, indicating thermalization. 

\begin{figure}
    \includegraphics[width=0.4\textwidth]{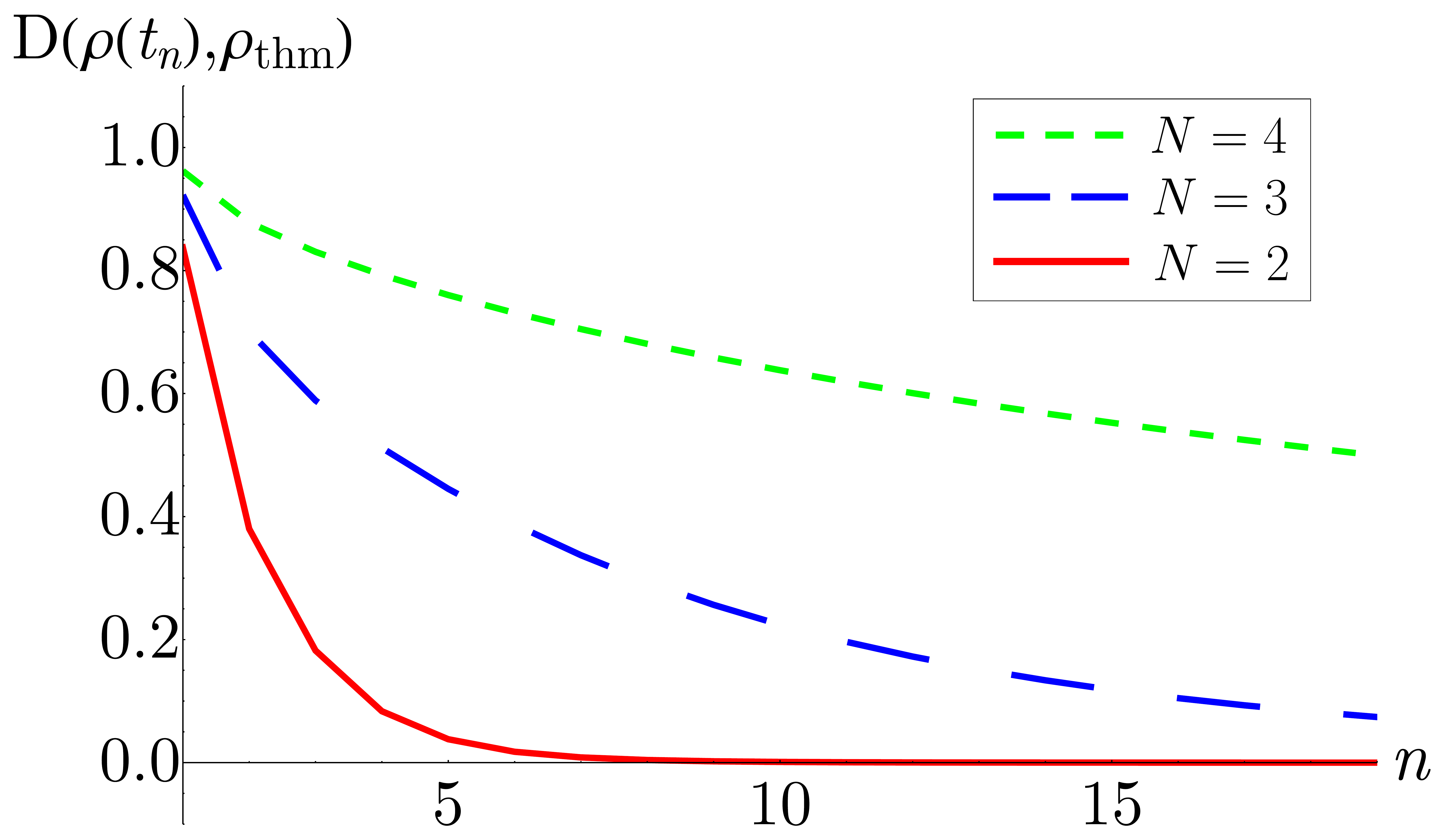}
    \caption{Trace distance between the time-dependent density matrix and the system thermal state density matrix as a function of the collisional model time step for the one-dimensional XXZ model with $N=2$ (red, solid), $N=3$ (blue, dashed), and $N=4$ (green, dotted). Parameters are $J = h = \Delta = 1$, $t_s = \Delta t = 1$, $g = 1$, and $\beta = 2$.
    }
    \label{fig:TDplot}
\end{figure}

To verify that thermalization also occurs at larger system sizes, we repeat our collisional model simulation for chains of length $N=3$ and $N=4$. As plotting each eigenstate occupation probability rapidly becomes unwieldy at larger system sizes, we instead use the trace distance as our measure of thermalization. The trace distance is defined as,
\begin{equation}
    D(\rho,\sigma) = \frac{1}{2} \tr{\sqrt{ (\rho - \sigma)^{\dagger}(\rho - \sigma)}}
\end{equation}
In Fig. \ref{fig:TDplot} we plot the trace distance between the time-dependent density matrix and the system thermal state density matrix as a function of the collisional model time step for the one-dimensional XXZ model. We see that the trace distance approaches zero as $n$ increases, demonstrating that the time-dependent density matrix converges to the thermal state. However, as system size increases, more collisions are required to thermalize the system. 

\section{Thermalization under the Monte Carlo Metropolis algorithm}
\label{sec:3}

Next, we construct a Monte Carlo algorithm with Metropolis \cite{metropolis1953} updating to demonstrate thermalization in the same context as the collisional model. The Metropolis updating technique \cite{metropolis1953} was introduced as an algorithm to significantly improve Monte Carlo convergence. The goal of the Metropolis scheme is to generate a sequence of states such that the distribution of these states closely resembles the desired distribution. The key feature of this sampling procedure is using the Metropolis filter function when deciding to accept or reject a proposed move. There are two steps involved in this method. The first one is to propose a move from the present state $x$ to $x^{\prime}$, which is based on the conditional proposal probability $G(x^{\prime}|x)$. The next step involves the acceptance probability $A(x^{\prime}|x)$ which then determines the acceptance of the proposed move. The detailed balance condition requires, $A(x^{\prime}|x) G(x^{\prime}|x)P(x) = A(x|x^{\prime}) G(x|x^{\prime})P(x^{\prime})$, where $P(x)$ is the underlying probability distribution to be sampled. Note that the normalization factor in $P(x)$ gets canceled and thus the stochastic averaging can be done without explicitly computing the normalization factor, one key advantage of the Metropolis method. In the context of thermalization, $P(x)$ is the Gibbs distribution, with the partition function as the normalization factor. Thus the accept and reject criteria are based on the energy difference between the proposed and the current state. For classical systems, for instance, spin systems, the updates are often local, involving the flipping of the spin at the chosen site. However, in general, the update scheme can also be non-local \cite{Troyer2003}. Since the spin basis states are the eigenstates of the classical spin Hamiltonians, these ``classical" updates are sufficient to produce thermalization. 

In the context of quantum systems, the eigenvectors are, in general, non-trivial superpositions of the spin basis states. In this case, the classical spin updates are not sufficient to achieve thermalization. To do so, the update scheme must ensure the algorithm explores all the system eigenstates. This can be done using eigenstate jump operators, whose action produces a jump from the present eigenstate to any other eigenstate. The simplest choice for the proposal probability $G(x^{\prime}|x)$ is a uniform distribution, meaning that all possible ``jumps" from the present eigenstate to other eigenstates will be proposed with equal probability.

For studying thermalization, we consider an update scheme based on  jumps between different eigenstates. The steps of the algorithm are outlined in Algorithm~\ref{MetropolAlgo}. The first step involves computing the eigenvectors $\{\psi_i\}$ and eigenvalues $\{E_i\}$ of the XXZ Hamiltonian given in Eq.~\ref{hamil_xxz} for a system of size $N$. The set of eigenvectors and eigenvalues can then be used for the Metropolis updating scheme. We consider an arbitrary initial state, $\rho_0$, which can be expressed in the energy eigenbasis as $\rho_0 = \sum_{k,l} a_{k,l} \ket{\psi_k} \bra{\psi_l}$. We then propose a transition $A_{i,j} = \ket{j}\bra{i}$, with $j \neq i$, where $\ket{i}$ is an eigenstate with non-zero population in $\rho_0$ and $\ket{j}$ is chosen randomly. We then use the Metropolis condition based on the energy difference $E_j - E_i$ to decide whether the jump is accepted. If the jump is not accepted, we update the initial state to $\frac{1}{2}(\ket{i}\bra{i}\rho_0 + \rho_0 \ket{i}\bra{i})$. This process serves to kill off the populations and coherences in $\rho_0$ other than those associated with $\ket{i}$, and acts as a discrete Monte Carlo implementation of the Lindblad decoherence term \cite{BreuerPetruccione2007}. This step is then repeated many times, and the results are averaged over to construct the density matrix for the next time step. Repeating the entire sequence of proposal, accept/reject, and then averaging, $n$ times results in the construction of a time series that gives the time evolution of the density matrix. Note that across Monte Carlo runs for each time step coherences are only preserved by sequences of rejections and are reduced by half each time. Consequently, the coherences of the averaged time-dependent density matrix decrease exponentially, as expected in the decoherence process.     

Our approach is related to the circuit-based quantum Metropolis scheme proposed in Ref. \cite{Temme2011}. In this scheme a unitary operator is applied that rotates the initial eigenstate into a superposition state. If a transition is accepted, a measurement selects one element of this superposition state as the next eigenstate in the time series. Our Metropolis scheme takes a different approach by randomly choosing one out of all the potential eigenstate transitions, with equal probability given to each transition. Our algorithm also differs by directly incorporating decoherence during the rejection step. This allows states with non-zero coherences to be taken as an input to the algorithm, and (partially) preserves some coherences in the output time series. Other quantum implementations of Monte Carlo algorithms have also been proposed, such as the Monte-Carlo wave function method \cite{Dalibard1992, Molmer1993, Plenio1998} and the quantum trajectories method \cite{Daley2014} used in quantum optics.

This stochastic sampling protocol in the Metropolis algorithm is different conceptually from the collisional model approach. The microscopic modeling of the interaction and the partial tracing steps in the collisional models is replaced by proposing and then accepting or rejecting the jumps between the eigenstates. In other words, the system and the bath interaction followed by the partial tracing of the bath leads to the mixedness in the system density matrix in the collisional model picture. On other hand, in the Metropolis scheme the thermal density matrix is constructed from averaging over many runs. This has computational advantage, as the numerical steps involves dealing with Hilbert space of the system alone, while the same is not true in the collisional models. In the later approach, the combined system and the bath evolves under the interaction Hamiltonian, as mentioned in Eq.~\ref{eq:CMevolution}, thereby involving the computation over the combined Hilbert space.

\begin{algorithm*}
\caption{Thermalization using Metropolis Algorithm}
\SetKwInput{KwInput}{Input} 
\KwInput{System size $N$, Hamiltonian parameters $h_1, h_2, \ldots $, Inverse temperature $\beta$}
\KwInput{Number of thermalization steps $num\_thermalization$, Number of Monte Carlo runs $num\_runs$.}
\SetKwFunction{FMain}{Metropolis}
\SetKwProg{Fn}{Function}{:}{\KwRet}
\SetKwFunction{FSum}{Exact diag}
\Fn{\FSum{$h_1, h_2, \ldots$}}{
       . \\
       .\\ 
        \KwRet $\{\ket{\psi}\}, \{E\}$\;
}
\Fn{\FMain}{
Exact diag($h_1, h_2, \ldots $)\;
\KwData{Initial state $\rho_0$}
Initialize system into initial state $\rho_0$\;
\For{$n \leftarrow 1$ \KwTo $num\_thermalization$} {
\For{$m \leftarrow 1$ \KwTo $num\_runs$}{
    Propose jump $\ket{\psi'}\bra{\psi}$ where $\ket{\psi}$ is an eigenstate with non-zero population in $\rho$ and $\psi' \neq \psi$\;
    Calculate the energy of state $\psi$ as 
    $E = \langle \psi | H | \psi \rangle$, and similar 
    for state $\psi'$. Compute the energy difference $\Delta E = E^{'} - E$.
    Calculate the acceptance ratio $\alpha = \min\left(1, {\rm exp}(-\beta \Delta E) \right)$\;
    Generate a random number $u$ from a uniform distribution $[0, 1]$\;
    \eIf{$u < \alpha$}{
        Accept the proposed jump: apply jump operator $\ket{\psi'}\bra{\psi}\rho\ket{\psi}\bra{\psi'}$\;
    }{
        Reject the proposed jump: apply decoherence map $\frac{1}{2}(\ket{\psi}\bra{\psi}\rho + \rho \ket{\psi}\bra{\psi})$\;
    }
}
Average over Monte Carlo runs to generate current state $\rho'$.
}   
}
\label{MetropolAlgo}
\end{algorithm*}

\begin{figure}
\centering
    \includegraphics[width=0.4\textwidth]{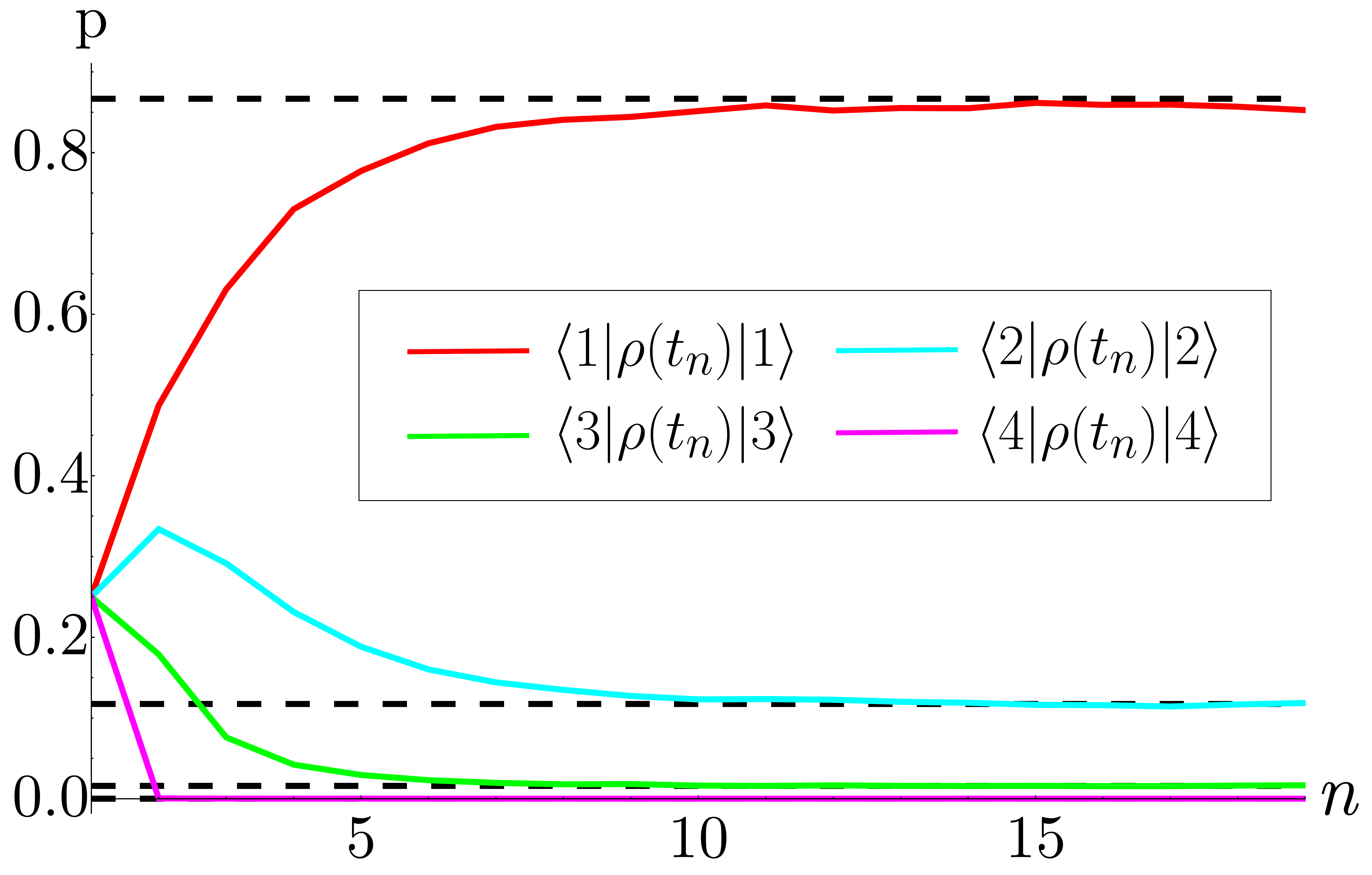}
    \caption{Occupation probabilities of each energy eigenstate of the two-site XXZ model as a function of the Metropolis algorithm time step, averaged over 100,000 runs. Horizontal black dotted lines indicate the thermal state occupation probabilities for each eigenstate. Parameters are $J = h = \Delta = 1$ and $\beta = 2$.} 
    \label{fig:thermal_metropol}    
\end{figure}

As an example, the thermalization of the XXZ model using the Metropolis technique is illustrated in Fig.~\ref{fig:thermal_metropol}.  The parameters in the Hamiltonian are: $J = 1, h = 1, \Delta = 1$, and the inverse temperature is $\beta =2$. As in the case of the collisional model, we have taken the initial state to be a pure state composed of an even superposition of all the energy eigenstates. We can see the probabilities of the evolving state quickly approach the expected thermal probability values based on the thermal density matrix at the chosen inverse temperature.   

\section{Equivalence of both models}
\label{sec:4}

In the previous two sections, we demonstrated that both the collisional model and Metropolis algorithm can result in thermalization. We note that, qualitatively, the evolution of the occupation probabilities appear very similar in both cases, as can be seen by comparing Figs. \ref{fig:ThermXXZCM} and \ref{fig:thermal_metropol}. In this section we prove explicitly that, under certain conditions, the dynamics generated by both models are identical by showing that the density matrices produced by each individual time step are identical.     

\subsection{Collisional model: Single time step evolution}

Let us first consider a single time step of the collisional model. The density operator after the interaction with the bath ancilla is given by,
\begin{equation}
    \label{eq:CMforComp}
    \rho_S(t_{n+1}) = \mathrm{tr}_{a}\{U_{I}(t_{n+1},t_n)\left[\rho_S(t_n) \otimes \rho_{a} \right]U_{I}^{\dagger}(t_{n+1},t_n)\}. 
\end{equation}
We note that in this equation we have not explicitly included the system evolution operators as in Eq.~\eqref{eq:CMevolution}. For the purposes of comparing the repeated interaction and Metropolis schemes we are only concerned with the dynamics generated during the system-environment interactions and thus any system evolution that occurs before a collision can be folded into the definition of $\rho_S(t_n)$. 

The interaction unitary in Eq.~\eqref{eq:CMforComp} can be expanded iteratively using the Dyson series,
\begin{equation}
    U_{I}(t_{n+1},t_n) = I - i \int_{t_n}^{t_{n+1}} ds \, H_I(s) U_I(s,t_0)
\end{equation}
Taking the limit that the interaction time becomes very short, we can truncate the Dyson series at the second order,
\begin{equation}
    \label{eq:Uexp}
    U_{I}(t_{n+1},t_n) \approx I - i \Delta t H_I -\frac{1}{2} \Delta t^2 H_I^2  
\end{equation}
where we have assumed that $H_I(t) = H_I$ is time independent and $\Delta t \equiv t_{n+1}-t_n$. Using this expression for the time evolution operator in Eq. \eqref{eq:CMforComp} we have,
\begin{equation}
    \label{eq:tevolrho}
    \begin{split}
    \rho_S&(t_{n+1}) = \mathrm{tr}_{a}\Big\{ \rho_S(t_n) \otimes \rho_{a} + i \Delta t \rho_S(t_n) \otimes \rho_{a} H_I \\& - i \Delta t H_I \rho_S(t_n) \otimes \rho_{a} - \frac{(\Delta t)^2}{2} \, H_I^2 \rho_S(t_n) \otimes \rho_{a} \\& - \frac{(\Delta t)^2}{2} \,  \rho_S(t_n) \otimes \rho_{a} H_I^2 + (\Delta t)^2 H_I \rho_S(t_n) \otimes \rho_{a} H_I \\& - \frac{i(\Delta t)^3}{2} \, H_I^2 \rho_S(t_n) \otimes \rho_{a} H_I+ \frac{i(\Delta t)^3}{2} \, H_I \rho_S(t_n) \otimes \rho_{a} H_I^2 \\& + \frac{(\Delta t)^4}{2} \, H_I^2 \rho_S(t_n) \otimes \rho_{a} H_I^2 \Big\}.   
    \end{split}
\end{equation}
The interaction Hamiltonian can be expressed in the general form,
\begin{equation}
    H_I = g \sum_{\alpha} \left(A_{\alpha} \otimes B_{\alpha}^{\dagger} + A_{\alpha}^{\dagger} \otimes B_{\alpha} \right)
\end{equation}
where $A_{\alpha}$ and $B_{\alpha}$ are operators in the Hilbert spaces of the system and bath, respectively. Under the assumption that $g\Delta t \ll 1$ we truncate Eq.~\eqref{eq:tevolrho} at the second order in $\Delta t$. We further note that any terms in Eq.~\eqref{eq:tevolrho} containing an odd power of $H_I$ will vanish. This is due to the fact that odd number bath correlation functions such as $\mathrm{tr}\{B_{\alpha} \rho_{a}\}$ and $\mathrm{tr}\{B_{\alpha}^{\dagger} \rho_{a}\}$ are zero. Thus we can simplify Eq.~\eqref{eq:tevolrho} to,
\begin{equation}
    \label{eq:tevolrhoSimp}
    \begin{split}
    \rho_S&(t_{n+1}) = \rho_S(t_n) + \Delta t^2 \mathrm{tr}_{a}\Big\{ H_I \rho_S(t_n) \otimes \rho_{a} H_I  \\&  - \frac{1}{2} H_I^2 \rho_S(t_n) \otimes \rho_{a} - \frac{1}{2} \rho_S(t_n) \otimes \rho_{a} H_I^2 \Big\}.
    \end{split}
\end{equation}

We next assume that the bath ancillae are all prepared in identical thermal states structured so that their spectra fulfill the transition energy matching condition, as described in section \ref{sec:2}. Plugging the interaction Hamiltonian from Eq.~\eqref{eq:InteractionH} and the bath density matrix from Eq.~\eqref{eq:BathTherm} into Eq. \eqref{eq:tevolrhoSimp} and carrying out the partial trace yields,
\begin{equation}
\label{eq:CMFinal}
\begin{split}
    \rho_S(t_{n+1})& = \rho_S(t_n) + \frac{(g \Delta t)^2}{Z_a} \sum_{i>j} \bigg[A_{i,j} \rho_S(t_n) A_{i,j}^{\dagger}  \\&  + A_{i,j}^{\dagger} \rho_S(t_n) A_{i,j}  e^{-\beta \epsilon_{\alpha_{i,j}}} - \frac{1}{2} \Big\{\rho_S(t_n), \ket{i}\bra{i} \Big\} \\& \qquad - \frac{e^{-\beta \epsilon_{\alpha_{i,j}}}}{2}\Big\{\rho_S(t_n), \ket{j}\bra{j} \Big\} \bigg],
\end{split}
\end{equation}
where $\left\{A,B\right\} \equiv AB + BA$ is the standard anticommutator. Note that, without loss of generality, we have also set the ground state energy of the bath at zero $\epsilon_0 = 0$. 

We pause for a moment here to review the assumptions we have made so far. Equation~\eqref{eq:CMFinal} provides a discrete master equation for the time evolution of the system density matrix generated by a single, very short duration interaction with a thermal bath ancilla. Furthermore, the spectrum of the bath ancilla is engineered such that there exists an energy gap between the ground state and an excited state of the bath for every possible energy eigenstate transition of the system. 

\subsection{Metropolis algorithm: Single time step evolution}

Now let us consider the same situation, namely how the average system state evolves under a single time step of the Metropolis algorithm. For a system with $d$ eigenstates, and thus $L = d-1$ possible transitions between an occupied energy eigenstate to another unoccupied eigenstate, the average system state after a transition is given by,
\begin{equation}
     \frac{1}{L} \sum_{i \neq j} \left[ A_{i,j} \rho_S(t_n) A_{i,j}^{\dagger} \right] 
\end{equation}
where, as in the case of the collisional model, $A_{i,j} = \ket{j}\bra{i}$ is the transition operator between system energy eigenstates $\ket{i}$ and $\ket{j}$. Under the Metropolis algorithm, a transition is accepted with probability $e^{-\beta \omega_{i,j}}$ where $\omega_{i,j}$ is defined as,
\begin{equation}
    \omega_{i,j} = \begin{cases}        0       & E_j - E_i < 0 \cr
                 E_j - E_i & E_j - E_i > 0 \end{cases}
\end{equation}
In the typical Metropolis scheme, if a transition is rejected, the state of the system is left unchanged. However, in order to properly incorporate decoherence we modify this procedure to instead apply the map $\mathcal{L}_i[\rho] = \frac{1}{2}(\ket{i}\bra{i}\rho + \rho \ket{i}\bra{i})$. Thus the average system state after a rejection is,
\begin{equation}
    \frac{1}{L} \sum_{i \neq j} \mathcal{L}_i[\rho_S(t_n)] = \frac{1}{2 L} \sum_{i \neq j} (\ket{i}\bra{i}\rho_S(t_n) + \rho_S(t_n) \ket{i}\bra{i}) 
\end{equation}

Accounting for both the accept and reject possibilities, the average system state after a single time step is,
\begin{equation}
\label{eq:MetroAvg}
\begin{split}
    \rho_S(t_{n+1}) &= \frac{1}{L} \sum_{i \neq j} \Big[ A_{i,j} \rho_S(t_n) A_{i,j}^{\dagger} e^{-\beta \omega_{i,j}} \\
    &+ \left(1 - e^{-\beta \omega_{i,j}}\right) \frac{1}{2} (\ket{i}\bra{i}\rho_S(t_n) + \rho_S(t_n) \ket{i}\bra{i})\Big]
\end{split}
\end{equation}

Equation \eqref{eq:MetroAvg} simplifies to,
\begin{equation}
\begin{split}
    \rho_S(t_{n+1}) = \rho_S(t_n) +& \frac{1}{L} \sum_{i \neq j} \Big[ \Big( A_{i,j} \rho_S(t_n) A_{i,j}^{\dagger} \\ &- \frac{1}{2} \Big\{\rho_S(t_n), \ket{i}\bra{i} \Big\}\Big)e^{-\beta \omega_{i,j}} \Big]
\end{split}
\end{equation}

In order to account for the piece-wise structure of $\omega_{i,j}$ we separate the double summation into terms where $i<j$ and $i>j$,
\begin{equation}
\label{eq:Metroij}
\begin{split}
    \rho_S(t_{n+1}) &= \rho_S(t_n) + \frac{1}{L} \sum_{i>j} \Big[ \Big( A_{i,j} \rho_S(t_n) A_{i,j}^{\dagger} \\&- \frac{1}{2} \Big\{\rho_S(t_n), \ket{i}\bra{i} \Big\} \Big] + \frac{1}{L} \sum_{i<j} \Big[ \Big( A_{i,j} \rho_S(t_n) A_{i,j}^{\dagger} \\& \qquad - \frac{1}{2} \Big\{\rho_S(t_n), \ket{i}\bra{i} \Big\}\Big)e^{-\beta (E_j - E_i)} \Big]
\end{split}
\end{equation}
Noting that $A_{i,j} = A_{j,i}^{\dagger}$ we can swap the indices in the second summation of Eq.~\eqref{eq:Metroij} rewrite it as a single summation,
\begin{equation}
\label{eq:MetroFinal}
\begin{split}
\rho_S(t_{n+1})& = \rho_S(t_n) + \frac{1}{L} \sum_{i>j} \bigg[A_{i,j} \rho_S(t_n) A_{i,j}^{\dagger}  \\&  + A_{i,j}^{\dagger} \rho_S(t_n) A_{i,j}  e^{-\beta (E_i - E_j)} - \frac{1}{2} \Big\{\rho_S(t_n), \ket{i}\bra{i} \Big\} \\& \qquad - \frac{e^{-\beta (E_i - E_j)}}{2}\Big\{\rho_S(t_n), \ket{j}\bra{j} \Big\} \bigg].
\end{split}
\end{equation}
Recalling that, in the collisional model, we structured the bath energies such that $E_i - E_j = \epsilon_{\alpha_{i,j}}$ we can compare Eqs. \eqref{eq:MetroFinal} and \eqref{eq:CMFinal} and see that they are identical when the condition $1/L = (g \Delta t)^2/Z_a$ is satisfied. Implicit in this equivalence is the premise that the duration of the time step, $t_{n+1}-t_n$, is the same for both the Metropolis algorithm and the collisional model. As the Metropolis algorithm is a purely phenomenological model, the time step duration is simply a model parameter. Thus we are free to choose it be identical to the physically-motivated time interval used in the collisional model, as discussed in Section \ref{sec:2}.

\subsection{Conditions of equivalence}

In this subsection we will consider in detail the conditions necessary to fulfill the equivalence demonstrated in the previous two subsections. Analytically, we have shown that the discrete time evolution of the density matrix generated by the collisional model, Eq.~\eqref{eq:CMFinal}, and the Metropolis algorithm, Eq.~\eqref{eq:MetroFinal} have exactly the same structure. Notably, this equivalence does not explicitly depend on a particular system size or choice of model Hamiltonian. However, we note that the truncated expansion in Eq.~\eqref{eq:tevolrhoSimp} used to derive the time-evolved collisional model density matrix is only accurate under the condition that $g \Delta t \ll 1$. Furthermore, the collisional model bath ancillae must be structured in a thermal state that fulfills the transition energy matching condition. Finally, as mentioned previously, the exact equivalence between Eqs. \eqref{eq:MetroFinal} and \eqref{eq:CMFinal} requires $1/L = (g \Delta t)^2/Z_a$. 

\begin{figure}
    \includegraphics[width=0.4\textwidth]{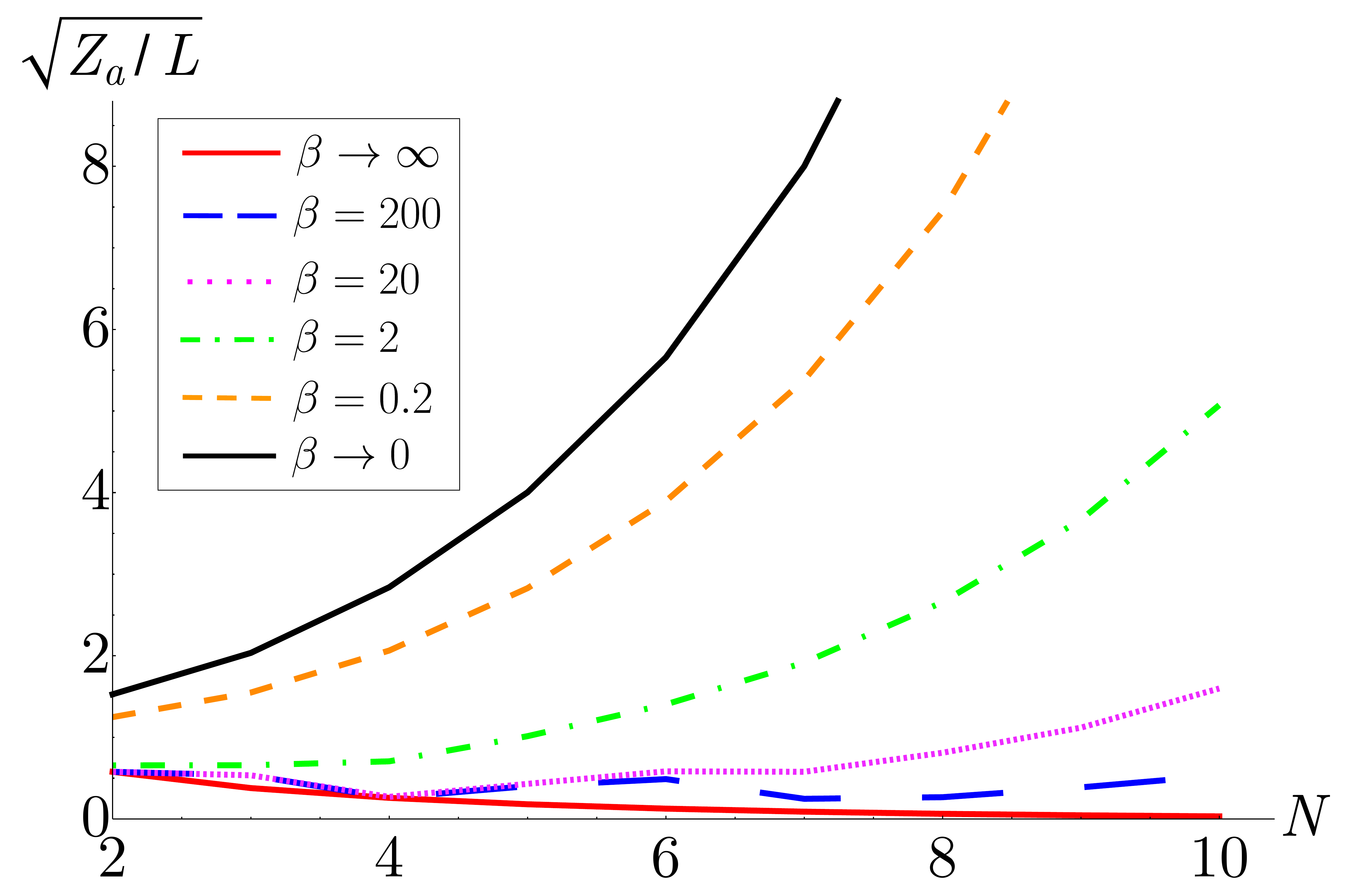}
    \caption{Ratio of the partition function for the collisional model bath ancillae, $Z_a$, to the number of possible eigenstate transitions at each time step in the Metropolis algorithm, $L$, for the one-dimensional XXZ model as a function of chain length $N$ at inverse temperature $\beta = 200$ (blue, long dashed), $\beta = 20$ (pink, dotted), $\beta = 2$ (green, dot-dashed), $\beta = 0.2$ (orange, short dashed), and $\beta \rightarrow 0$ (black, solid). The optimal zero temperature lower bound for the case of a non-degenerate spectrum is included for comparison (red, solid). Parameters are $J = h = \Delta = 1$.}
    \label{fig:ZBLplot}
\end{figure}

In order to satisfy both the conditions $g \Delta t \ll 1$ as well as $1/L = (g \Delta t)^2/Z_a$ we see that we want the ratio of $Z_a/L$ to be as small as possible. In Fig. \ref{fig:ZBLplot} we plot this ratio as a function of chain length for different temperatures. At low temperatures, the ratio remains relatively flat with system size, while at high temperatures it grows exponentially with system size.

\begin{figure}
    \includegraphics[width=0.4\textwidth]{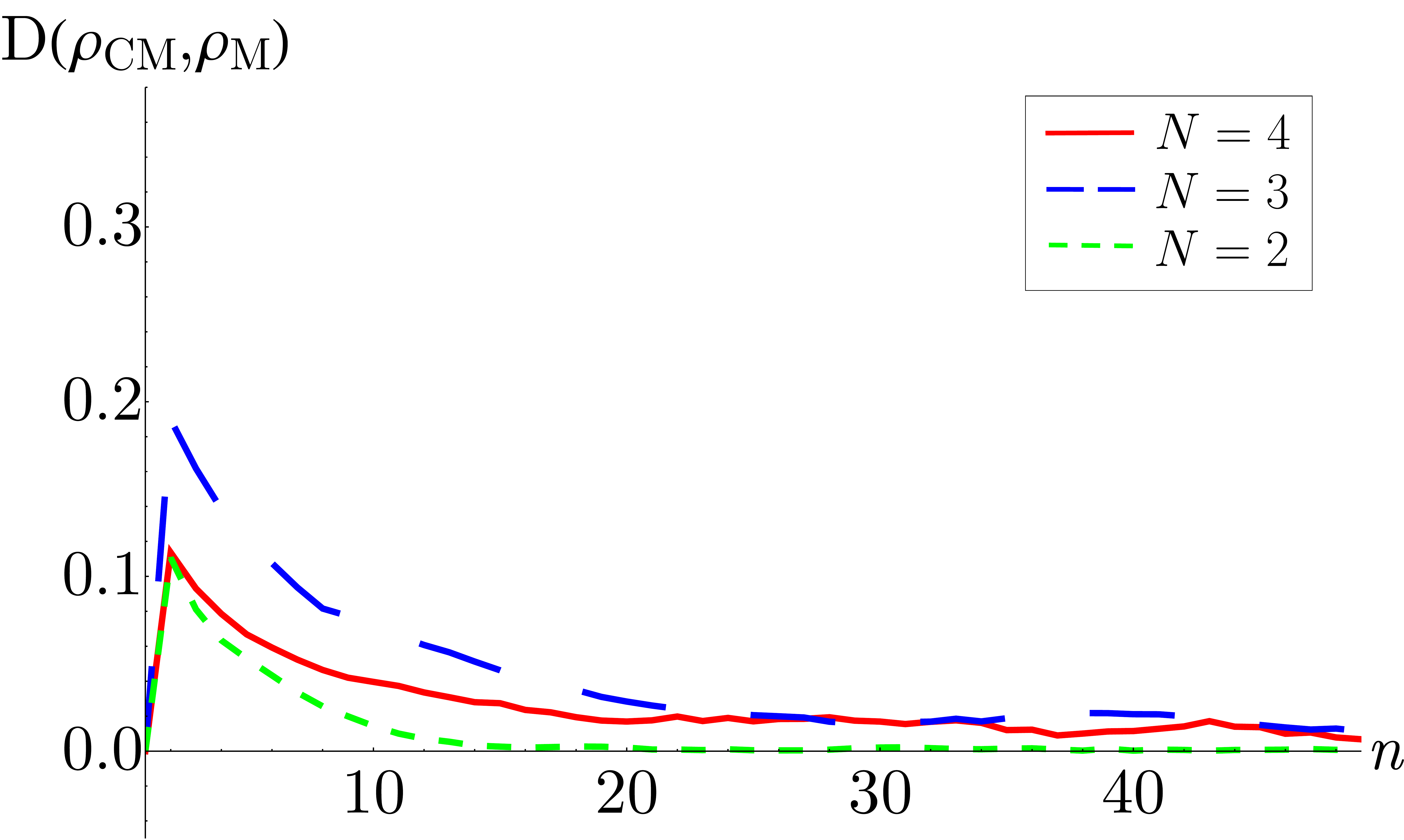}
    \caption{Trace distance between the density matrix arising from the collisional model and the density matrix arising from the Metropolis algorithm as a function of model time steps, $n$, for a one-dimensional XXZ chain of length $N = 4$ (red, solid), $N=3$ (blue, long dashed), and $N = 2$ (green, short dashed). Parameters are $J = h = \Delta = 1$ and $\beta = 20$. For the collisional model we have fixed the interaction parameter $g \Delta t = \sqrt{Z_a/L}$ and for the Metropolis algorithm we have constructed the density matrix from averaging over $10^6$ runs.}
    \label{fig:TDMCMplot}
\end{figure}

This behavior can be understood by considering the high and low temperature limits of $Z_a$. We recall that the bath partition function is given by $Z_a = \sum_{\alpha_{i,j}} e^{-\beta \epsilon_{\alpha_{i,j}}}$, where $\epsilon_{\alpha_{i,j}}$ corresponds to the magnitude of the energy difference between system energy eigenvalues $E_i$ and $E_j$. For an $N$-site spin chain system the number of eigenvalues is $2^N$. In this case, we have $L = 2^N - 1$ while the upper limit of the summation in $Z_a$ will be $\binom{2^N}{2} + 1$. Thus, in the infinite temperature limit we have,
\begin{equation}
    \lim_{\beta \rightarrow 0} \frac{Z_a}{L} = \frac{\binom{2^N}{2}+1}{2^N - 1} = \frac{2^N}{2} + \frac{1}{2^N-1} 
\end{equation}
Here we see that as the system size grows, the ratio $Z_a/L$, and thus also $g \Delta t$ grow, rendering the approximate expansion for the time evolution operator increasingly inaccurate and leading to the different dynamics between the models observed in Fig. \ref{fig:TDMCMplot}.

On the other hand, in the zero temperature limit, the only terms that contribute to $Z_a$ are those where $\epsilon_{\alpha_{i,j}} = 0$. This occurs in the case of the ground state energy of the bath and for any $\epsilon_{\alpha_{i,j}}$ corresponding to degenerate pairs of system energy eigenvalues. As the number of degenerate energy states depends on the system size and model parameters, this leads to the non-monotonic behavior of $Z_a/L$ observed at low temperatures in Fig. \ref{fig:ZBLplot}. In the optimal scenario, where there are no degenerate eigenvalues, the ratio becomes,
\begin{equation}
    \lim_{\beta \rightarrow \infty} \frac{Z_a}{L} = \frac{1}{2^N - 1}
\end{equation} 
In general, as long as the number of pairs of degenerate eigenvalues grows slower than the total number of eigenvalues, the ratio $Z_a/L$ will remain small at low temperatures.

To compare the validity of the equivalence, in Fig. \ref{fig:TDMCMplot} we plot the trace distance between the density matrix generated by Eq.~\eqref{eq:CMforComp} for the collisional model and the density matrix arising from averaging over $10^6$ runs of the Metropolis scheme for the one-dimensional XXZ model. Specifically, we use Eq.~\eqref{eq:CMforComp} for the collisional model, rather than Eq.~\eqref{eq:CMFinal}, as we wish to test the regimes in which the approximations that went into the derivation of Eq.~\eqref{eq:CMFinal} are valid. In our comparison, we fix the collisional model interaction parameter using the condition $(g \Delta t)^2= Z_a/L$. Thus, the only source of the deviation between the density matrices for both models comes from the fact that the truncated expansion of the time evolution operator does not fully capture the dynamics of the collisional model. We see that the trace distance first increases (initially it is zero, as both methods start from the same initial state), before decreasing, as both models ultimately result in the thermal state density matrix. 

We also see that the amount by which the trace distance increases depends on the system size. The increase is the largest for a system size of $N=3$, but as system size is increased further to $N=4$ the initial increase in trace distance falls back to the level of the $N=2$ case. This behavior is partially explained by Fig.~\ref{fig:ZBLplot} where we see that, for an inverse temperature of $\beta=20$, the ratio $Z_a/L$ decreases between $N=2$ and $N=4$, thus rendering the second order approximation in Eq.~\eqref{eq:Uexp} increasingly more accurate.  The larger trace distance for $N=3$, despite similar values of $Z_a/L$, occurs due to the fact that the trace distance is also affected by the choice of initial state, which impacts the rank of the matrix $\rho_{\mathrm{M}} - \rho_{\mathrm{CM}}$ at subsequent time steps. The choice of an initial state that is a pure state composed of an even superposition of all energy eigenstates produces the maximum rank for $\rho_{\mathrm{M}} - \rho_{\mathrm{CM}}$ in the initial dynamics. However, even in this case, we see that the contribution to the trace distance from $Z_a/L$ is much more significant, as evidenced by the subsequent decrease in trace distance for $N=4$, which has a lower value of $Z_a/L$.

In summary, the equivalence demonstrated here holds best at low temperatures and large system sizes of models that have relatively few degeneracies in their spectrum. The accuracy at large system sizes is particularly relevant, as these are precisely the regimes where the collisional model dynamics are most computationally expensive. Notably, while the calculation of the partition function $Z_a$ is computationally expensive, the partition function is not actually used in the Metropolis algorithm. It is only necessary to verify the equivalence. If there is a priori reason to believe a model has a well spaced spectrum with few degeneracies, and thus that $\sqrt{Z_a/L}$ is small, then $Z_a$ is not required to generate the dynamics from the Metropolis algorithm.

\section{Concluding Remarks} 
\label{sec:5}

In this work we have verified that both the collisional model framework and the Metropolis algorithm lead to thermalization of a many-body system when, in the case of the collisional model, the spectrum of the bath ancillae corresponds to each of the energy eigenstate transitions in the system. We have then demonstrated analytically that not only do both schemes produce thermalization, but that the time-dependent dynamics generated by both models are exactly equivalent when a condition relating the collisional model interaction strength and the ratio of bath ancilla partition function to number of possible energy eigenstate transitions is fulfilled. 

The Metropolis algorithm is typically considered to be a purely phenomenological method of sampling possible system configurations, while collisional models are constructed to simulate underlying microscopic interactions that give rise to the system dynamics. These results demonstrate that, assuming the equivalence conditions are met, the Metropolis algorithm can, in fact, capture the same dynamics of the thermalization process as the collisional model. This could make the Metropolis model useful in situations where the transient dynamics leading up to thermalization are important, such as studying the performance of finite time quantum heat engines \cite{Feldmann1996, Kloc2019, Das2020} or verifying stochastic thermodynamic behavior such as fluctuation theorems \cite{Funo2018}.

Despite the similarities discussed here, there remain some important distinctions between the collisional model and the Metropolis algorithm. By microscopically modeling the system-bath interaction, the collisional model is significantly more general and can be applied with arbitrarily structured bath ancillae to study non-thermal steady states of open system dynamics.

In terms of computational cost, to simulate thermalizing dynamics both models require computing the energy eigenstates of the system, which is a computationally expensive process that scales exponentially with the system size. However, the Metropolis algorithm is advantaged in that it simply tracks individual eigenstate transitions and needs to know only about the Hilbert space of the system. The collisional model must take into account the much larger joint system-bath Hilbert space and perform repeated partial traces after each system-bath interaction. Thus the cost of the collisional model grows exponentially with the combined system plus bath ancilla dimensions, while the cost of the Metropolis scheme grows exponentially with the system size alone.

For example, consider a system whose Hilbert space dimension is $d$. For modeling thermalizing dynamics using a bath whose spectrum has an excited state corresponding to each transition in the system, as we have done in our demonstration of thermalization in the XXZ model, the Hilbert space of each bath ancilla must be of dimension $\binom{d}{2}+1$. Thus the total Hilbert space that the collisional model must operate in is of dimension $\mathcal{O}(d^3)$. Even in the most optimal scenario, when the spectra of the bath ancillae are identical to the spectrum of the system, the joint system bath Hilbert space will be of dimension $\mathcal{O}(d^2)$. We must also keep in mind that the Metropolis scheme does incur an additional cost from the fact that the dynamics must be averaged over many repeated runs in order to construct the time-dependent density matrix, while the the collisional model directly models the density matrix evolution. However, each Metropolis experimental run still consists of only a series of products of $d$-dimensional operators.  

These results open several avenues for potential future work. Here we have shown an equivalence between the collisional model and Metropolis approaches in thermalizing dynamics for short collisional model interaction times and subject to the condition that the ratio of bath partition function to number of system energy eigenstate transitions is small. It may be possible to generalize these results to generic open system dynamics by modifying the distribution that the accept/reject probabilities are drawn in the Monte Carlo approach. Collisional models have seen extensive use in studying non-Markovian open system dynamics. It would be interesting to see if a similar equivalence could be found between non-Markovian collisional models and non-Markovian Monte Carlo schemes for quantum evolution \cite{Breuer1999, Maniscalco2004, Piilo2005}. 

\begin{acknowledgments}

We acknowledge support from AFOSR (FA9550-23-1-0034,FA2386-21-1-4081) and ARO (W911NF2210247). We thank the HPC resources at Virginia Tech, where some of the results in this manuscript are generated. 

\end{acknowledgments}
\hfill \break
\appendix

\onecolumngrid

\twocolumngrid

\bibliography{CollisionalMetropolis}
		
\end{document}